
\def\singlespace{\normalbaselines}
\def\oneandahalfspace{\baselineskip=1.15\normalbaselineskip plus 1pt
\lineskip=2pt\lineskiplimit=1pt}

\def\np{\vfill\eject}

\def\nofirstpagenoten{\nopagenumbers\footline={\ifnum\pageno>1\tenrm
\hss\folio\hss\fi}}
\def\nofirstpagenotwelve{\nopagenumbers\footline={\ifnum\pageno>1\twelverm
\hss\folio\hss\fi}}
\def\leaderfill{\leaders\hbox to 1em{\hss.\hss}\hfill}
\def\ft#1#2{{\textstyle{{#1}\over{#2}}}}
\def\frac#1/#2{\leavevmode\kern.1em
\raise.5ex\hbox{\the\scriptfont0 #1}\kern-.1em/\kern-.15em
\lower.25ex\hbox{\the\scriptfont0 #2}}
\def\sfrac#1/#2{\leavevmode\kern.1em
\raise.5ex\hbox{\the\scriptscriptfont0 #1}\kern-.1em/\kern-.15em
\lower.25ex\hbox{\the\scriptscriptfont0 #2}}


\parindent=20pt
\def\narrow{\advance\leftskip by 40pt \advance\rightskip by 40pt}

\def\AB{\bigskip
        \centerline{\bf ABSTRACT}\medskip\narrow}
\def\nonarrower{\advance\leftskip by -40pt\advance\rightskip by -40pt}
\def\AE{\bigskip\nonarrower}

\def\boxit#1{\vbox{\hrule\hbox{\vrule\kern3pt
        \vbox{\kern3pt#1\kern3pt}\kern3pt\vrule}\hrule}}

\def\gtorder{\mathrel{\raise.3ex\hbox{$>$}\mkern-14mu
             \lower0.6ex\hbox{$\sim$}}}
\def\ltorder{\mathrel{\raise.3ex\hbox{$<$}|mkern-14mu
             \lower0.6ex\hbox{\sim$}}}
\def\dalemb#1#2{{\vbox{\hrule height .#2pt
        \hbox{\vrule width.#2pt height#1pt \kern#1pt
                \vrule width.#2pt}
        \hrule height.#2pt}}}

\font\fourteentt=cmtt10 scaled \magstep2
\font\fourteenbf=cmbx12 scaled \magstep1
\font\fourteenrm=cmr12 scaled \magstep1
\font\fourteeni=cmmi12 scaled \magstep1
\font\fourteenss=cmss12 scaled \magstep1
\font\fourteensy=cmsy10 scaled \magstep2
\font\fourteensl=cmsl12 scaled \magstep1
\font\fourteenex=cmex10 scaled \magstep2
\font\fourteenit=cmti12 scaled \magstep1
\font\twelvett=cmtt10 scaled \magstep1 \font\twelvebf=cmbx12
\font\twelverm=cmr12 \font\twelvei=cmmi12
\font\twelvess=cmss12 \font\twelvesy=cmsy10 scaled \magstep1
\font\twelvesl=cmsl12 \font\twelveex=cmex10 scaled \magstep1
\font\twelveit=cmti12
\font\tenss=cmss10
 
 \font\ninebf=cmbx7 scaled \magstep1
\font\ninerm=cmr7 scaled \magstep1 \font\ninei=cmmi7 scaled \magstep1
\font\ninesy=cmsy7 scaled \magstep1 
\font\eightrm=cmr7 scaled 1140 
 
\font\sevenbf=cmbx7 \font\sevenrm=cmr7 \font\seveni=cmmi7
\font\sevensy=cmsy7 

\catcode`@=11
\newskip\ttglue
\newfam\ssfam

\def\fourteenpoint{\def\rm{\fam0\fourteenrm}
\textfont0=\fourteenrm \scriptfont0=\tenrm \scriptscriptfont0=\sevenrm
\textfont1=\fourteeni \scriptfont1=\teni \scriptscriptfont1=\seveni
\textfont2=\fourteensy \scriptfont2=\tensy \scriptscriptfont2=\sevensy
\textfont3=\fourteenex \scriptfont3=\fourteenex \scriptscriptfont3=\fourteenex
\def\it{\fam\itfam\fourteenit} \textfont\itfam=\fourteenit
\def\sl{\fam\slfam\fourteensl} \textfont\slfam=\fourteensl
\def\bf{\fam\bffam\fourteenbf} \textfont\bffam=\fourteenbf
\scriptfont\bffam=\tenbf \scriptscriptfont\bffam=\sevenbf
\def\tt{\fam\ttfam\fourteentt} \textfont\ttfam=\fourteentt
\def\ss{\fam\ssfam\fourteenss} \textfont\ssfam=\fourteenss
\tt \ttglue=.5em plus .25em minus .15em
\normalbaselineskip=16pt
\abovedisplayskip=16pt plus 4pt minus 12pt
\belowdisplayskip=16pt plus 4pt minus 12pt
\abovedisplayshortskip=0pt plus 4pt
\belowdisplayshortskip=9pt plus 4pt minus 6pt
\parskip=5pt plus 1.5pt
\setbox\strutbox=\hbox{\vrule height12pt depth5pt width0pt}
\let\sc=\tenrm
\let\big=\fourteenbig \normalbaselines\rm}
\def\fourteenbig#1{{\hbox{$\left#1\vbox to12pt{}\right.\n@space$}}}

\def\twelvepoint{\def\rm{\fam0\twelverm}
\textfont0=\twelverm \scriptfont0=\ninerm \scriptscriptfont0=\sevenrm
\textfont1=\twelvei \scriptfont1=\ninei \scriptscriptfont1=\seveni
\textfont2=\twelvesy \scriptfont2=\ninesy \scriptscriptfont2=\sevensy
\textfont3=\twelveex \scriptfont3=\twelveex \scriptscriptfont3=\twelveex
\def\it{\fam\itfam\twelveit} \textfont\itfam=\twelveit
\def\sl{\fam\slfam\twelvesl} \textfont\slfam=\twelvesl
\def\bf{\fam\bffam\twelvebf} \textfont\bffam=\twelvebf
\scriptfont\bffam=\ninebf \scriptscriptfont\bffam=\sevenbf
\def\tt{\fam\ttfam\twelvett} \textfont\ttfam=\twelvett
\def\ss{\fam\ssfam\twelvess} \textfont\ssfam=\twelvess
\tt \ttglue=.5em plus .25em minus .15em
\normalbaselineskip=14pt
\abovedisplayskip=14pt plus 3pt minus 10pt
\belowdisplayskip=14pt plus 3pt minus 10pt
\abovedisplayshortskip=0pt plus 3pt
\belowdisplayshortskip=8pt plus 3pt minus 5pt
\parskip=3pt plus 1.5pt
\setbox\strutbox=\hbox{\vrule height10pt depth4pt width0pt}
\let\sc=\ninerm
\let\big=\twelvebig \normalbaselines\rm}
\def\twelvebig#1{{\hbox{$\left#1\vbox to10pt{}\right.\n@space$}}}

\def\tenpoint{\def\rm{\fam0\tenrm}
\textfont0=\tenrm \scriptfont0=\sevenrm \scriptscriptfont0=\fiverm
\textfont1=\teni \scriptfont1=\seveni \scriptscriptfont1=\fivei
\textfont2=\tensy \scriptfont2=\sevensy \scriptscriptfont2=\fivesy
\textfont3=\tenex \scriptfont3=\tenex \scriptscriptfont3=\tenex
\def\it{\fam\itfam\tenit} \textfont\itfam=\tenit
\def\sl{\fam\slfam\tensl} \textfont\slfam=\tensl
\def\bf{\fam\bffam\tenbf} \textfont\bffam=\tenbf
\scriptfont\bffam=\sevenbf \scriptscriptfont\bffam=\fivebf
\def\tt{\fam\ttfam\tentt} \textfont\ttfam=\tentt
\def\ss{\fam\ssfam\tenss} \textfont\ssfam=\tenss
\tt \ttglue=.5em plus .25em minus .15em
\normalbaselineskip=12pt
\abovedisplayskip=12pt plus 3pt minus 9pt
\belowdisplayskip=12pt plus 3pt minus 9pt
\abovedisplayshortskip=0pt plus 3pt
\belowdisplayshortskip=7pt plus 3pt minus 4pt
\parskip=0.0pt plus 1.0pt
\setbox\strutbox=\hbox{\vrule height8.5pt depth3.5pt width0pt}
\let\sc=\eightrm
\let\big=\tenbig \normalbaselines\rm}
\def\tenbig#1{{\hbox{$\left#1\vbox to8.5pt{}\right.\n@space$}}}
\let\rawfootnote=\footnote \def\footnote#1#2{{\rm\parskip=0pt\rawfootnote{#1}
{#2\hfill\vrule height 0pt depth 6pt width 0pt}}}

\def\tenfoot{\tenpoint\hskip-\parindent\hskip-.1cm}

\overfullrule=0pt
\twelvepoint
\oneandahalfspace
\def\sbullet{\raise.2em\hbox{$\scriptscriptstyle\bullet$}}
\nofirstpagenotwelve
\hsize=16.5 truecm
\baselineskip 15pt

\def\ket#1{\big|{#1}\big\rangle}
\def\ft#1#2{{\textstyle{{#1}\over{#2}}}}

\def\del{\partial}

\oneandahalfspace
\rightline{CTP TAMU--64/92}
\rightline{Imperial/TP/91-92/40}
\rightline{hep-th/9209111}
\rightline{September 1992}

\vskip 2truecm
\centerline{\bf Discrete States in the $W_3$ String}
\vskip 1.5truecm
\centerline{C.N. Pope$^{(1)}$\footnote{$^*$}{\tenfoot \sl  Supported in
part by the U.S. Department of Energy, under
grant DE-FG05-91ER40633.}, E. Sezgin$^{(1)}$\footnote{$^\diamond$}{\tenfoot
\sl   Supported in part by the National Science Foundation, under grant
PHY-9106593.},  K.S. Stelle$^{(2)}$\footnote{$^\$$}{\tenfoot \sl Supported in
part by the Commission of the European Communities under Contract
SC1*-CT91-0674.} and X.J. Wang$^{(1)}$}

\vskip 1.5truecm
\centerline{\it $^{(1)}$ Center
for Theoretical Physics,
Texas A\&M University,}
\centerline{\it College Station, TX 77843--4242, USA.}
\bigskip

\centerline{\it ${(2)}$ The Blackett Laboratory, Imperial College,}
\centerline{\it \phantom{xxxxxxx} Prince Consort Road, London SW7 2BZ, UK.}

\vskip 1.5truecm
\AB\singlespace
   We construct the low-lying discrete states of the two-scalar $W_3$
string. This includes states that correspond to the analogues of the ground
ring generators of the ordinary two-dimensional string. These give rise to
infinite towers of discrete states at higher levels.
\AE\oneandahalfspace

\np
\noindent
{\bf 1. Introduction}
\bigskip

     A two-dimensional string, having no transverse dimensions, might
naively be expected to have no physical content, beyond the tachyon.
However, as is now well known, there are in fact discrete excited states at
special values of the momenta [1].  These states occur with non-standard ghost
numbers as well as the standard one [2,3].  The physical operators
associated with these states give rise to an infinite-dimensional algebra of
conserved charges [3].  The Ward identity associated with this symmetry
provides a powerful tool for determining the correlation functions of the
theory.

     It is natural to investigate the analogous questions for the $W_3$
string.  There seems to be no notion of a critical dimension for a $W_3$
string, since a critical realisation of the $W_3$ algebra in terms of scalar
fields always requires background charges, regardless of the number of
scalars.  To this extent the $W_3$ string is like the so-called
``non-critical string,'' namely a string where criticality is achieved by
choosing $d\ne 26$ scalar fields, with background charges.  The special
feature of $d=2$ for ordinary strings is that the gauge symmetries of the
excited states remove all transverse degrees of freedom.  It is not so clear
for the $W_3$ string what the analogous dimension should be.  Results from
previous work on $W$ strings, and the work described here, suggest that it
is most natural to study the case of a two-scalar realisation of $W_3$.  In
some sense, as we shall see, this theory exhibits some features of the $d=1$
string as well as some features of the $d=2$ string.

      In this paper, we shall construct some of the low-lying discrete
states of the two-scalar $W_3$ string, including some which exhibit properties
analogous to the ground ring discovered in [3].  In order to highlight the
similarities and the differences, we shall begin in section 2 by summarising
the salient features of the discrete states of the ordinary string.  In view
of the remark made above, it is appropriate to consider discrete states for
the one-dimensional string as well as the two-dimensional string. In section
3 we then describe the construction of discrete states in the $W_3$ string.
In particular, we find analogues of the $x$ and $y$ operators of Witten,
which are the building blocks of the ground ring.  Section 4 contains
conclusions and discussion.

\bigskip
\noindent{\bf 2. Discrete states in ordinary strings}
\bigskip

     With the exception of the tachyon, all physical states of the
two-dimensional string occur for only discrete values of the momenta.  By
definition, a physical state is one that is annihilated by the BRST operator
$Q_{BRST}$, whilst not itself being expressible as $Q_{BRST}$ of any state.
Physical states are characterised by their level number $\ell$ and their ghost
number $G$.  We use the convention in which $G=0$ for the standard ghost
vacuum $\ket{-}\equiv c_1\ket{0}$, where $\ket{0}$ is the
$SL(2,C)$-invariant vacuum.  A general state is obtained by acting on the
tachyon state $\ket{\vec p,-}$ with a string of ghost operators $c_{-m}, \
m\ge 0$, antighost operators $ b_{-m},\ m\ge 1$, and matter operators
$\vec\alpha_{-m},\ m\ge 1$. Thus one can easily see that the states at level
$\ell$ can have ghost numbers in the interval
$$
\Big[ {{1-\sqrt{8\ell +1}}\over 2}\Big]\le G\le
\Big[ {{1+\sqrt{8\ell +1}}\over 2}\Big]\ ,\eqno(2.1)
$$
where $[x]$ denotes the integer part of $x$.

For the two dimensional string, the physical discrete states at a
given level fall into quartets with the following ghost numbers: $(G,\ G+1,
\ G+1,\ G+2)$ and their conjugates with ghost numbers $(-G+1,\ -G,\ -G,\
-G-1)$. (The conjugate of a state with ghost number $G$ and momentum $(p_1,
p_2)$ is a state with ghost number $(-G+1)$ and momentum $(-p_1-2Q, -p_2)$.)
Starting from the state with ghost number $G$, which we shall call
the ``prime state,''  the remaining three
states of the quartet can be obtained by acting with $a_\varphi,\ a_X$ and
$a_\varphi a_X$, where $a_\varphi\equiv[Q_{BRST}, \varphi]=c\del\varphi
-\sqrt2 \del c$ and $a_X\equiv[Q_{BRST},  X]=c\del X$.  Here $\varphi$ is
the Liouville field with background charge  $Q=\sqrt{2}$ and $X$ is the matter
field.\footnote{$^\dagger$}{\tenfoot An  alternative description of the
mappings induced by $a_\varphi$ and $a_X$ is  as follows.  Consider the
prime state $\ket{\vec p_0,G}$, with discrete momentum $\vec p_0$,
satisfying $Q_{BRST}\ket{\vec p_0,G}=0$. Extrapolating  this to arbitrary
momentum in the exponential gives a state $\ket{\vec p,G}$ that is
annihilated by $Q_{BRST}$ at $\vec p=\vec p_0$.  Since for generic $\vec p$
the state $Q_{BRST}\ket{\vec p,G}$ is manifestly BRST invariant, it follows
that the states  $$
\ket{\vec p,G+1}_i\equiv{d\over d p_i} Q_{BRST}\ket{\vec p,G}\Big|_{\vec p
=\vec p_0}, \qquad i=1,2
$$
are also annihilated by $Q_{BRST}$.  However they are not BRST trivial,
since at $\vec p=\vec p_0$ they cannot be written as $Q_{BRST}$ acting on
any states. (This is a restatement of an argument in [3].)
They are in fact the same as
the states obtained by acting on $\ket{\vec p_0,G}$ with $a_\varphi$ or
$a_X$.  To see this, note that $\ket{\vec p,G}$ can be written as
$R(b,c,\del\vec\varphi)e^{\vec p\cdot\vec\varphi(0)}\ket{-}$, and so
${d\over d p_i} Q_{BRST}\ket{\vec p,G}\Big|_{\vec p =\vec
p_0}=[Q_{BRST},\varphi_i]R(b,c,\del\vec\varphi)e^{\vec p_0\cdot\vec\varphi}
(0)\ket{-}$, where $\vec\varphi=(\varphi,X)$.}  The operators $a_\varphi$ [4]
and $a_X$ are both BRST non-trivial, even though they are formally written
as BRST commutators, since neither $\varphi$ nor $X$ is a well-defined
conformal field.  The action of the $a_\varphi$ and $a_X$ operators is
indicated in Fig.\ 1 below.
$$
\matrix{ &  & \ket{G} &  &  \cr
 &  a_\varphi\swarrow & &  \searrow a_X &  \cr
\cr
&\ket{G+1}_1   &  &   \ket{G+1}_2 &\cr
\cr
  & a_X\searrow  & & \swarrow a_\varphi & \cr
 &  & \ket{G+2} &  &  \cr\cr }\qquad \qquad
\matrix{ &  & \ket{-G-1} &  &  \cr
 &  a_\varphi\swarrow & &  \searrow a_X &  \cr
\cr
&\ket{-G}_1   &  &   \ket{-G}_2 &\cr
\cr
  & a_X\searrow  & & \swarrow a_\varphi & \cr
 &  & \ket{-G+1} &  &  \cr\cr }
$$
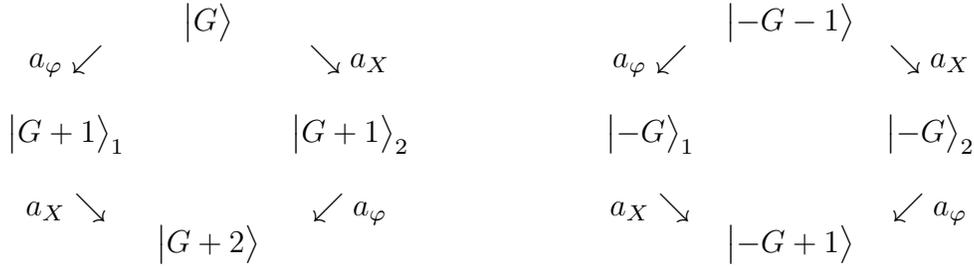
\centerline{\sl Figure 1}
{\tenpoint\sl \phantom{zzzzz} The structure of the states in a quartet and
its conjugate, generated from a prime state $\ket{G}$.}
\bigskip
\bigskip

       At level $\ell=0$, we have the tachyon $\ket{\vec p,-}$, with $G=0$.
This is an exception to the above rules, firstly because it has
continuous momentum and secondly because it is annihilated by $a_{X}$.
Thus the generic quartet structure degenerates in this case. Acting with
$a_\varphi$ on  $\ket{\vec p,-}$ gives the $G=1$ state $\ket{\vec p,+}$.
This is in fact the conjugate of the tachyon $\ket{\vec p',-}$, with $\vec
p'=-\vec p-(2Q,0)$.

     At level $\ell=1$, there is a single discrete state at the lowest ghost
number, $G=-1$.  This is $b_{-1}\ket{\vec 0,-}$; in other words the
$SL(2,C)$ vacuum $\ket{0}$, with zero momentum.  Acting with the $a_\varphi$
and $a_X$ operators then fills out a quartet, with ghost numbers
$(-1,0,0,1)$.  The conjugates of these states give a quartet with ghost
numbers $(2,1,1,0)$. The prime state $\ket{0}$ corresponds to the $G=0$
identity operator.\footnote{$^\dagger$}{\tenfoot Note that the quartet
structure that we have been describing is equivalent to the ``diamond''
structure of [4].  However, in our description we use the two BRST non-trivial
operators $a_\varphi$ and $a_X$, rather than just the $a_\varphi$ in [4],
giving us explicit generators for all the states in the quartets.  States
built with $a_X$ may differ from those in [3,4] by BRST-trivial states.  For
example, at level $\ell=1$ the conjugate state $\ket{-G}_2$ with $G=1$ has
the form $c_{-1}\ket{\vec p,-}$, whereas the corresponding state given in [3]
is of the form $\alpha^1_{-1}\ket{\vec p,+}$ (where $\alpha^1_n$ denotes the
Fourier modes of the Liouville field $\varphi$).  The two differ by the
BRST-trivial state $Q_{BRST}b_{-1}\ket{\vec p,+}$.}

    At level $\ell=2$, there are two discrete states at the lowest ghost
number, $G=-1$. Each of these prime states has its quartet partners and
the conjugate quartet.  The operators corresponding to these two prime
states are the $x$ and $y$ ground ring generators of Witten, with $G=0$.
They are given by
$$
\eqalignno{
x&=\Big(b\, c +\ft12Q(\del\varphi +i\del X)\Big) e^{\ft12Q
(\varphi-i X)},&(2.2a)\cr
y&=\Big(b\, c +\ft12Q(\del\varphi -i\del X)\Big) e^{\ft12Q
(\varphi+i X)},&(2.2b)\cr}
$$
where the background-charge parameter $Q$ takes the value $\sqrt2$.
The low-lying states, up to level 2, are depicted in Fig.\ 2 below.

\bigskip\bigskip\bigskip
{\tenpoint \settabs 3\columns
\+ \phantom{$G=-1$} \phantom{$G=-1$}Level 0& \phantom{$G=-1$} \phantom{$G=-1$}
Level 1& \phantom{$G=-1$} \phantom{$G=-1$} Level
2\cr \+&&\cr
\+$G=-1$
&$\quad\quad$---------$\bullet$------------------------&$
\quad\quad$---------$\bullet$------------------------\cr
\+ \phantom{$G=-1$}
&$\qquad\quad\ \ \, \swarrow\quad \searrow$
  &$\qquad\quad\ \  \swarrow\quad \searrow$\cr
\+$G=\phantom{-}0\quad$
---------$\bullet$---------$\circ$---------&
$\quad\quad$---$\bullet$-----------$\bullet$---------$\circ$---------&
$\quad\quad$---$\bullet$-----------$\bullet$---------$\circ$---------\cr
\+ \phantom{$G=-1\quad$}
$ \qquad\swarrow \qquad\ \swarrow$
&$\qquad\quad\ \ \,\searrow\quad\swarrow
\qquad\ \ \swarrow\quad\searrow$
&$\qquad\quad\ \ \,\searrow\quad\swarrow
\qquad\ \ \swarrow\quad\searrow$  \cr
\+$G=\phantom{-}1\quad$
---$\bullet$---------$\circ$---------------&$\quad\quad$
---------$\bullet$---------$\circ$-----------$\circ$---&$\quad\quad$
---------$\bullet$---------$\circ$-----------$\circ$---\cr
\+ \phantom{$G=-1\quad$}
&$\qquad\qquad\qquad\qquad\quad\ \,\searrow\quad\swarrow$
&$\qquad\qquad\qquad\qquad\quad\ \,\searrow\quad\swarrow$\cr
\+$G=\phantom{-}2$
&$\quad\quad$--------------------------$\circ$---------
&$\quad\quad$--------------------------$\circ$---------\cr
}
\bigskip
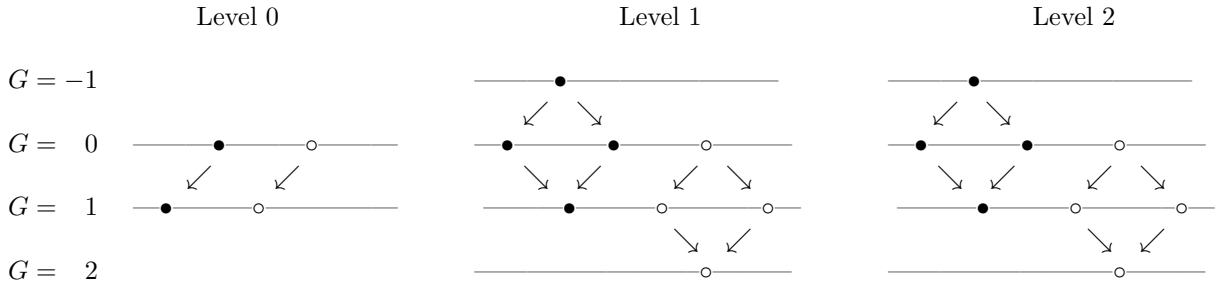
\centerline{\sl Figure 2}
{\tenpoint\sl
This diagram shows the structure of quartets and their
conjugates for each prime state, at levels 0, 1 and 2 in the two-scalar string.
Horizontal lines indicate permitted ghost numbers, bullets indicate the quartet
states, and circles indicate their conjugates.  At level 0 we have a
continuum of tachyon prime states; the diamonds degenerate to dumbells in
this case.  At level 1, we have the $SL(2,C)$ vacuum as prime state.  At level
2, there are two prime states, corresponding to the $x$ and $y$ ring
generators. }
\bigskip\bigskip

   Note that all the prime states discussed so far are manifestly
cohomologically non-trivial, since they each  have the lowest possible ghost
number for their level and so there are no states of which they can be the
BRST variations. Conversely, the duals of these prime states, having maximum
possible ghost numbers for their levels, are manifestly BRST invariant. All
the higher-level prime states correspond to operators that are monomials
of $x$ and $y$ [3].

    As we mentioned in the introduction, in some respects the two-scalar $W_3$
string resembles the ordinary string with one scalar. Accordingly it is
useful to examine the single-scalar string in more detail. The single scalar
$\varphi$ is a Liouville field, with energy-momentum tensor $T=-\ft12
(\del\varphi )^2-Q\del^2\varphi$, where the background charge is given by
$Q^2=\ft{25}{12} $. Since now we only have one scalar, there is just one
multiplet-generating operator $a_\varphi=[Q_{BRST},\varphi]=c\del\varphi-
Q\del c$. Hence the multiplets are doublets rather than quartets, at all
levels.

     For the single-scalar string the tachyon states
$\ket{p,-}=e^{p\varphi(0)}\ket{-}$ become discrete also, with momenta
$p_+=-\ft65 Q$ and $p_-=-\ft45 Q$. At level $\ell=1$, there is the $SL(2,C)$
vacuum, its doublet partner, and their conjugates. At level $\ell=2$, there
is now just a single prime state, at $G=-1$, given by the operator
$$
x=(b\, c+\ft35 Q\del\varphi) e^{\ft25 Q\varphi}  \eqno(2.3)
$$
acting on the $SL(2,C)$ vacuum.  At this point a significant difference
between the one-scalar string and the two-scalar string emerges.  For the
two-scalar string, the exponential operators in the expressions for the ring
generators $x$ and $y$ are of the form $\exp\big[{1\over\sqrt2}(\varphi
\pm iX)\big]$ [3] (see eq.\ (2.2$a$-$b$), with $Q=\sqrt2$).  Consequently in
the OPE of $x$ with $x$ or $y$ with $y$ the  exponentials contribute no
$(z-w)$ factor, and in the OPE of $x$ with $y$  they contribute $(z-w)^{-1}$.
In the one-scalar case, on the other hand, we  see from (2.3) that in the OPE
of
$x$ with $x$ the exponentials will  contribute a factor of $(z-w)^{-1/3}$.  It
follows that only monomials of  the form
$$
x^n,\qquad {\rm for}\qquad n=3p\quad {\rm or}\quad n=3p+1,\eqno(2.4)
$$
where $p$ is a non-negative integer, give operators corresponding to states of
well-defined level number $\ell$.  The level number is given by
$$
\ell=\ft16 (n+2)(n+3)\ .\eqno(2.5)
$$
The first few allowed levels are $\ell=1,2,5,7,12,15,\cdots$, corresponding
to the operators $1,x,x^3,x^4,x^6,x^7,\cdots$.  The operators
$\{x^{3p}\}=1,x^3,x^6,x^9,x^{12},\cdots$ form a ring, generated by $x^3$.
The remaining operators $\{x^{3p+1}\}$ are then obtained as $x$ multiplied
by powers of the ring generator.

     An important property of the operator $x$ in (2.3) is that it
maps the tachyon $\ket{-\ft65 Q,-}$ into the tachyon $\ket{-\ft45 Q,-}$.
Specifically, the normal-ordered product of $x$ with the $p=-\ft65 Q$ tachyon
operator gives the $p=-\ft45 Q$ tachyon operator.  It is by seeking the
appropriate generalisation of this property that we find the clue, in the next
section, to the level number at which the analogous ring generators arise in
the two-scalar $W_3$ string.

\np

\noindent{\bf 3. Discrete states in the $W_3$ string}
\bigskip

     As we discussed in the introduction, we shall concentrate mainly on the
two-scalar realisation of $W_3$ [5].  The spin-2 and spin-3 primary currents
are given by
$$
\eqalignno{
T&=-\ft12(\del\varphi_1)^2 -\ft12(\del\varphi_2)^2 -Q_1 \del^2\varphi_1 -Q_2
\del^2 \varphi_2\ ,&(3.1a)\cr
W&={2i\over 3\sqrt{29}}\Big[-\ft13(\del\varphi_1)^3 + \del\varphi_1(\del
\varphi_2)^2 -Q_1 \del\varphi_1\del^2\varphi_1 +2Q_2\del \varphi_1 \del^2
\varphi_2\cr
&\qquad\qquad  +Q_1 \del\varphi_2 \del^2\varphi_2 -\ft13 Q_1^2 \del^3 \varphi_1
+Q_1Q_2 \del^3 \varphi_2\Big]\ ,  &(3.1b)\cr}
$$
where $Q_1$ and $Q_2$ are chosen such that
$$
Q_1^2=\ft{49}8,\qquad\qquad Q_2^2=\ft{49}{24}\ ,  \eqno(3.2)
$$
in order that the central charge take its critical value $c=100$.

     The BRST operator has the form
$$
Q_{BRST}=\oint dz \Big(c(T+\ft12 T_{\rm gh})+\gamma(W+\ft12
W_{\rm gh})\Big)\ . \eqno(3.3)
$$
The ghost currents $T_{\rm gh}$ and $W_{\rm gh}$ are given by [6,7]
$$
\eqalignno{
T_{\rm gh}&=-2b\,\partial c-\partial
b\, c-3\beta\, \partial\gamma-2\partial\beta\, \gamma\ , &(3.4a)\cr
W_{\rm gh}&=-\partial\beta\,
c-3\beta\, \partial c-\ft8{261}\big[\partial(b\, \gamma\,  T)+b\,
\partial\gamma \, T\big]\cr
&\ \ +\ft{25}{6\cdot261}\hbar \Big(2\gamma\, \partial^3b+9\partial\gamma\,
\partial^2b +15\partial^2\gamma\,\partial b+10\partial^3\gamma\,
b\Big)\ ,&(3.4b)\cr}
$$
where the ghost-antighost pairs ($c$,$b$) and ($\gamma$,$\beta$) correspond
respectively to the $T$ and $W$ generators.

The standard ghost vacuum for the $W_3$ string is defined by acting on the
$SL(2,C)$ vacuum with $c_1\gamma_1\gamma_2$ to give
$$
\ket{- -}\equiv c_1\gamma_1\gamma_2 \ket{0}\ .  \eqno(3.5)
$$
Tachyon states are constructed as $\ket{\vec p,- -}\equiv e^{\vec p\cdot
\vec\varphi(0)}\ket{- -}$. General states are obtained by acting on $\ket{\vec
p,- -}$ with a string of ghost operators $c_{-m}$, $\gamma_{-m}$, $m\ge 0$,
antighost operators $b_{-m}$, $\beta_{-m}$, $m\ge 1$, and matter operators
$\vec\alpha_{-m}$, $m\ge 1$. Thus the ghost number at level $\ell$ must lie
in the interval
$$
\Big[ 1-\sqrt{4\ell +1}\Big]\le G\le
\Big[ 1+\sqrt{4\ell +1}\Big]\ ,\eqno(3.6)
$$
where $[x]$ denotes the integer part of $x$.

     As for the ordinary string, the physical states are defined to be the
cohomologically nontrivial states annihilated by $Q_{BRST}$. We again have
multiplet-generating operators, this time given by $a_i=[Q_{BRST},
\varphi_i]$, $i=1,2$. They take the form
$$
\eqalign{
a_1 &=c\del\varphi_1-Q_1\del c+\ft8{261}\del\varphi_1 b\gamma\del\gamma
      -\ft8{261}Q_1(\del b\gamma\del\gamma+b\gamma\del^2\gamma)\cr
&\qquad+\ft{2i}{\sqrt{261}}\Big((\del\varphi_1)^2\gamma-Q_1\del\varphi_1\del
\gamma -(\del\varphi_2)^2\gamma-2Q_2\del^2\varphi_2\gamma
 +\ft13 Q_1^2\del^2\gamma \Big)  \cr
a_2 &=c\del\varphi_2-Q_2\del c+\ft8{261}\del\varphi_2 b\gamma\del\gamma
      -\ft8{261}Q_2(\del b\gamma\del\gamma+b\gamma\del^2\gamma)\cr
&\qquad+\ft{2i}{\sqrt{261}}\Big(-2\del\varphi_1\del\varphi_2\gamma
+2Q_2\del\varphi_1\del\gamma+2Q_2\del^2\varphi_1\gamma
+Q_1\del\varphi_2\del\gamma-Q_1Q_2\del^2\gamma\Big)\ . \cr}\eqno(3.7)
$$
Acting with $a_1$, $a_2$ and $a_1a_2$ on a prime state will generate
a quartet of states with ghost numbers $(G,\ G+1,\ G+1,\ G+2)$ and a
quartet of conjugates, which have ghost numbers $(-G+2,\ -G+1,\ -G+1,\-G)$.
(The conjugate of a state with ghost number $G$ and momentum $(p_1, p_2)$ is
a state with ghost number $(-G+2)$ and momentum $(-p_1-2Q_1, -p_2-2Q_2)$.)

     We now consider the physical states that are analogous to the level 0,
1 and 2 physical states of the ordinary string. At level zero we again have
tachyons. Like the one-scalar string, the tachyons in the two-scalar $W_3$
string have discrete momenta. This is because we have two constraints and
two momentum components. Since these constraints constitute a quadratic
polynomial (from $T$) and a cubic polynomial (from $W$), there are $2\times
3=6$ possible momentum values, which turn out to be:
$$
    \eqalign{
     \vec p_1 &=(-\ft67 Q_1, -\ft67 Q_2),\qquad\qquad
     \vec p_2  =(-\ft87 Q_1, -\ft87 Q_2)\ , \cr
     \vec p_3 &=(-\ft87 Q_1, -\ft67 Q_2),\qquad\qquad
     \vec p_4  =(-\ft67 Q_1, -\ft87 Q_2)\ , \cr
     \vec p_5 &=(-Q_1, -\ft57 Q_2),\qquad\qquad\quad
     \vec p_6  =(-Q_1, -\ft97 Q_2)\ . \cr}\eqno(3.8)
$$
Note that the pairs of momenta on each line are conjugate to each other.
The six states $\ket{\vec p_i,--}$, $i=1,\ldots,6$ are prime states, which
give rise to six quartets {\it via} the action of the multiplet-generating
operators $a_1$ and $a_2$.  In fact these quartets, which each span the full
range of allowed ghost numbers ($0\le G \le 2$) at level 0, are pairwise
conjugate.

    In the ordinary string, the next level ($\ell=1$) has the $SL(2,C)$
vacuum $b_{-1}\ket{\vec 0,-}$ as prime state.  For the $W_3$ string, the
$SL(2,C)$ vacuum is also a prime state, but it now occurs at level $\ell=4$,
since it is written in terms of the ghost vacuum $\ket{\vec 0,- -}$ as
$b_{-1}\beta_{-1} \beta_{-2}\ket{\vec 0,- -}$.  We see from this that it has
ghost number $-3$, which is the lowest value allowed at level $\ell=4$.  It
gives rise to a quartet with ghost numbers $(-3,-2,-2,-1)$, and a conjugate
quartet with ghost numbers $(5,4,4,3)$ and momentum $(-2Q_1,-2Q_2)$.

     The question now arises as to the level number at which the analogues
of the ring generators $x$ and $y$ of the ordinary string will occur.
Clearly they should lie at a higher level than that of the $SL(2,C)$ vacuum,
which corresponds to the identity operator in the ring.  For the ordinary
string, they in fact occurred at one level higher, namely $\ell=2$.  For the
$W_3$ string it is not {\it a priori} obvious whether they should again occur
at one level higher than the $SL(2,C)$ vacuum ({\it i.e.} $\ell=5$), or
whether on the other hand they should occur at an even higher level.  To
resolve this, we recall the property noted at the end of section 2 for the
operator $x$ in the one-scalar string.  There, we observed that $x$ maps the
tachyon $\ket{-\ft65 Q,-}$ into the tachyon $\ket{-\ft45 Q,-}$.

     Since in the present case the tachyons also have discrete momenta,
$\vec p_i$ given by (3.8), it is natural to seek operators $x$, $y,\cdots$
that map one tachyon into another.  Thus the requirements for such operators
are first of all that they have momenta given by $\vec p_i -\vec p_j$ for
some tachyon momenta $\vec p_i$ and $\vec p_j$.  Secondly, in order that
such an operator $x$ can act on a tachyon to give another state with
well-defined level number, it must be that the OPE of the exponential factor
in $x$ with the tachyon must give an integer power of $(z-w)$.  Thus we must
have
$$
(\vec p_i-\vec p_j)\cdot \vec p_j ={\rm integer}.\eqno(3.9)
$$
By enumerating all possible such cases, we find that just two momentum
values are possible for the $x$-type operators, namely
$$
\vec p =(\ft27 Q_1,0),\quad {\rm and}\quad \vec p=(\ft17 Q_1, \ft37
Q_2).\eqno(3.10)
$$
Each of these momentum values gives an exponential operator with conformal
weight $\Delta=-\ft12 \vec p^2 -\vec p\cdot \vec Q= -2$.  The ghost vacuum,
$\ket{\vec 0,--}\equiv c_1\gamma_1\gamma_2 \ket{0}$ has conformal weight
$-4$, and so for the $x$ operators to have spin 0 (as they must, in order to
map tachyons to tachyons), they must have level $\ell=6$.

     We have learnt that the $x$-type operators in the $W_3$ string occur at
level $\ell=6$.  Of course these operators, which map tachyons to tachyons,
must have ghost number $G=0$.  This means that the corresponding {\it
states} at level 6 will have ghost number $G=-3$.  From (3.6), we see that
the lowest allowed ghost number at level 6 is $G=-4$.  Thus we have a rather
different situation from the ordinary string, in that the $x$-type state
does not have the minimum allowed ghost number.  This means that when we
solve for such states by demanding that they are annihilated by $Q_{BRST}$, we
must then check that they are not merely BRST variations of some $G=-4$
states.  Also, it raises the possibility that the $x$-type states might in
principle be non-prime states obtained by acting on a prime state of ghost
number $G=-4$ with $a_1$ and $a_2$.  In fact, this turns out not to be the
case, as we shall describe below.

     The considerations above lead us to look for spin 0, $G=0$, level 6
operators.  The most general such operator, at arbitrary momentum $\vec p$,
has 30 parameters $g_i$ for the prefactors of the exponential, and is given
in the Appendix in eq.\ $(A.2)$.  Some representative terms are $(\del^2\vec
\varphi,\ \del b c,\ \del\beta \gamma,\cdots )e^{\vec p\cdot \vec \varphi}$.
We then require that this operator be
annihilated by the BRST operator, given by (3.1-3.4).  The calculations are
straightforward but somewhat tedious, and are best performed by computer.
The condition of $BRST$ invariance leads to 186 linear, homogeneous
equations, with $\vec p$-dependent coefficients, for the 30 parameters
$g_i$.  At generic values of momentum $\vec p$, there is a unique solution,
up to overall scale.  This can be easily
understood, since there is a unique level 6 state with ghost number $G=-4$,
namely $b_{-1}b_{-2}\beta_{-1} \beta_{-2}\ket{\vec p,- -}$.  Acting on this
with $Q_{BRST}$ gives a state that is annihilated by $Q_{BRST}$, and must
therefore be the one mentioned above.  It is, of course, BRST trivial.

     At certain special values of the momentum, the 186 equations for the 30
coefficients $g_i$ turn out to leave two parameters, rather than just one,
undetermined.  In other words, there are two independent solutions at these
particular momentum values.  One is merely the specialisation of the generic
BRST-trivial state mentioned above, whilst the other is a new, BRST
non-trivial state.  There are in fact two values of the momentum $\vec p$
for which these non-trivial solutions occur, namely the two values given in
(3.10).  Thus we have independently arrived at the same two momentum values
that were indicated by the tachyon-mapping argument.  We shall denote the
two operators by $x$ and $y$:
$$
\eqalign{
x&=R_x \exp(\ft27 Q_1 \varphi_1),\cr
y&=R_y \exp(\ft17 Q_1 \varphi_1 +\ft37 Q_2 \varphi_2),\cr}\eqno(3.11)
$$
where the prefactors $R_x$ and $R_y$ are given in the Appendix, in eqs
$(A.2-A.4)$.  The operators (3.11) are the $W_3$ analogues of the $x$ and $y$
operators of the ordinary two-dimensional string.

     The fact that $x$ and $y$ map certain tachyon states to other tachyon
states gives an independent proof that they are BRST non-trivial. This
follows from the fact that the tachyon states themselves are BRST
non-trivial, and so, for example, if two such states $\ket{t_1}$ and
$\ket{t_2}$ are related by $\ket{t_2}=x\ket{t_1}$, then it cannot be that
$x=\{Q_{BRST},U\}$ for any operator $U$, since then one would have
$\ket{t_2}=Q_{BRST}U\ket{t_1}$, contradicting the BRST non-triviality of
$\ket{t_2}$.

     We also asserted above that $x$ and $y$ themselves correspond to prime
states, rather than being obtained by acting with $a_1$ or $a_2$ on $G=-4$
level-6 prime states.  We see this by establishing that there are no values
of the momentum $\vec p$ for which the (unique) $G=-4$ state $b_{-1} b_{-2}
\beta_{-1}\beta_{-2}\ket{\vec p,--}$ is annihilated by $Q_{BRST}$, and hence
there are no $G=-4$ prime states at this level.

     Just as for the $x$ operator in the one-scalar string, here, we cannot
consider the set of all monomials $x^m y^n$ for arbitrary integers $m$ and
$n$, since the operator products are not always well defined.  The signal
for a product's being well defined is that the momentum of the resulting
operator should be such as to yield an integer-weight exponential operator.
This ensures that the monomial $x^m y^n$ is single valued, and has a
well-defined level number.  From (3.11), it follows that the conformal
weight of the exponential factor in $x^m y^n$ is
$$
\Delta=-\ft14(m^2+n^2+mn+7m+7n),\eqno(3.12)
$$
and the level number of the corresponding state is
$$
\ell=4-\Delta=\ft14(m^2+n^2+mn+7m+7n+16)\ .\eqno(3.13)
$$
The allowed values of $(m,n)$ such that $\Delta$ is an integer are
$$
\Big\{(0,0),\ (0,1),\ (1,0),\ (1,2),\ (2,1),\ (2,2)\Big\}\ {\rm mod}\ 4\ .
\eqno(3.14)
$$
The allowed operators are thus all of the form
$$
x^{4p}y^{4q} \big\{1,\, x,\, y,\, xy^2,\, x^2y,\, x^2y^2\big\}\ ,\eqno(3.15)
$$
where $p$ and $q$ are arbitrary non-negative integers.  The operators $x^{4p}
y^{4q}$ form a ring, generated by $x^4$ and $y^4$.  There is in fact a
larger ring, generated by $x^4$, $y^4$ and $x^2 y^2$, but it is not so
straightforward to state the rules for generating all the operators in
(3.15) in terms of this enlarged ring.

     So far, we have identified the tachyon operators, at level $\ell=0$,
the identity operator, at $\ell=4$, and the $x$ and $y$ analogues of the
ring generators, at $\ell=6$.  We have a possibility here for a richer
structure than the ordinary string, where the tachyon, identity, and ring
operators occurred at levels 0, 1 and 2 respectively.  In particular, we can
now look for BRST non-trivial operators with level numbers in between those
of the tachyon, the identity, and the $x$ and $y$ operators.  We have found
that there are six such operators at level $\ell=1$.  They take the form
$$
(c\gamma \pm{i\over\sqrt{522}} \gamma\del\gamma)e^{\vec p\cdot\vec\varphi}\,
\eqno(3.16)
$$
with momenta given by
$$
    \eqalign{
     \vec p_1 &=(-\ft47 Q_1, -\ft27 Q_2),\qquad\qquad
     \vec p_2  =(-\ft47 Q_1, -\ft{12}7 Q_2)\ , \cr
     \vec p_3 &=(-\ft37 Q_1, -\ft57 Q_2),\qquad\qquad
     \vec p_4  =(-\ft37 Q_1, -\ft97 Q_2)\ , \cr
     \vec p_5 &=(-\ft67Q_1,0),\qquad\qquad\qquad
     \vec p_6  =(-\ft87Q_1,0)\ . \cr}\eqno(3.17)
$$
In (3.16), the $+$ sign is chosen for the momenta $\vec p_1$, $\vec p_2$ and
$\vec p_5$, and the $-$ sign for $\vec p_3$, $\vec p_4$ and $\vec p_6$.
The six operators correspond to states $(\beta_{-1}\mp{i\over
\sqrt{522}}b_{-1}) \ket{ \vec p,--}$, which have $G=-1$, the lowest allowed
ghost number at level 1. These prime states give rise to quartets and
conjugates in the usual way.

     We have not looked exhaustively at all allowed ghost numbers for levels
2, 3 and 5.  However, we have verified that there are no discrete states at
the lowest permitted ghost numbers for these levels.  The low-lying states
are depicted in Fig.\ 3 below.
\np

{\tenpoint \settabs 5\columns

\+ &\qquad Level 0 &\qquad Level 1 &\qquad Level 4 &\qquad Level 6\cr

\+ & & & &\cr

\+$G=-4$& & & & ------------------ \cr

\+ & & & & \cr

\+$G=-3$& & &--------$\bullet$--------- &--------$\bullet$---------\cr

\+ & & &$\quad\ \ \swarrow\ \searrow$ &$\quad\ \ \swarrow\ \searrow$
\cr

\+$G=-2$& & &---$\bullet$--------$\bullet$-----&---$\bullet$--------$
\bullet$-----\cr

\+ & & &$\quad\ \ \searrow\ \swarrow$ &$\quad\ \ \searrow\ \swarrow$
\cr

\+$G=-1$& &--------$\bullet$------------ &--------$\bullet$---------
&--------$\bullet$---------\cr

\+ & &$\quad\ \ \swarrow\ \searrow$ & & \cr

\+$G=\phantom{-}0$ &-----$\bullet$------------$\circ$-----
&---$\bullet$--------$\bullet$--------&------------------
&------------------\cr

\+ &$\ \ \swarrow\ \searrow\qquad \swarrow\ \searrow$
&$\quad\ \ \searrow\ \swarrow$ & & \cr

\+$G=\phantom{-}1$ &$\bullet$--------$\bullet$---$\circ$--------$\circ$
&--------$\bullet$---$\circ$--------$\ \cdots$
&------------------$\ \ \cdots$
&------------------\cr

\+ &$\ \ \searrow\ \swarrow\qquad\searrow\ \swarrow$
&$\qquad\ \ \ \,\swarrow\ \searrow$ & & \cr

\+$G=\phantom{-}2$ &-----$\bullet$------------$\circ$-----
&--------$\circ$--------$\circ$---&------------------ &------------------\cr

\+ & &$\qquad\ \ \, \ \searrow\ \swarrow$ & & \cr

\+$G=\phantom{-}3$ & &------------$\circ$--------
&---------$\circ$--------
&---------$\circ$--------\cr

\+ & & &$\qquad \swarrow\ \searrow$ &$\qquad \swarrow\ \searrow$  \cr

\+$G=\phantom{-}4$ & & &-----$\circ$--------$\circ$---
&-----$\circ$--------$\circ$---\cr

\+ & & &$\qquad \searrow\ \swarrow$ &$\qquad \searrow\ \swarrow$  \cr

\+$G=\phantom{-}5$ & & &---------$\circ$--------
&---------$\circ$--------\cr

\+ & & & & \cr

\+$G=\phantom{-}6$ & & & &------------------ \cr
}
\bigskip
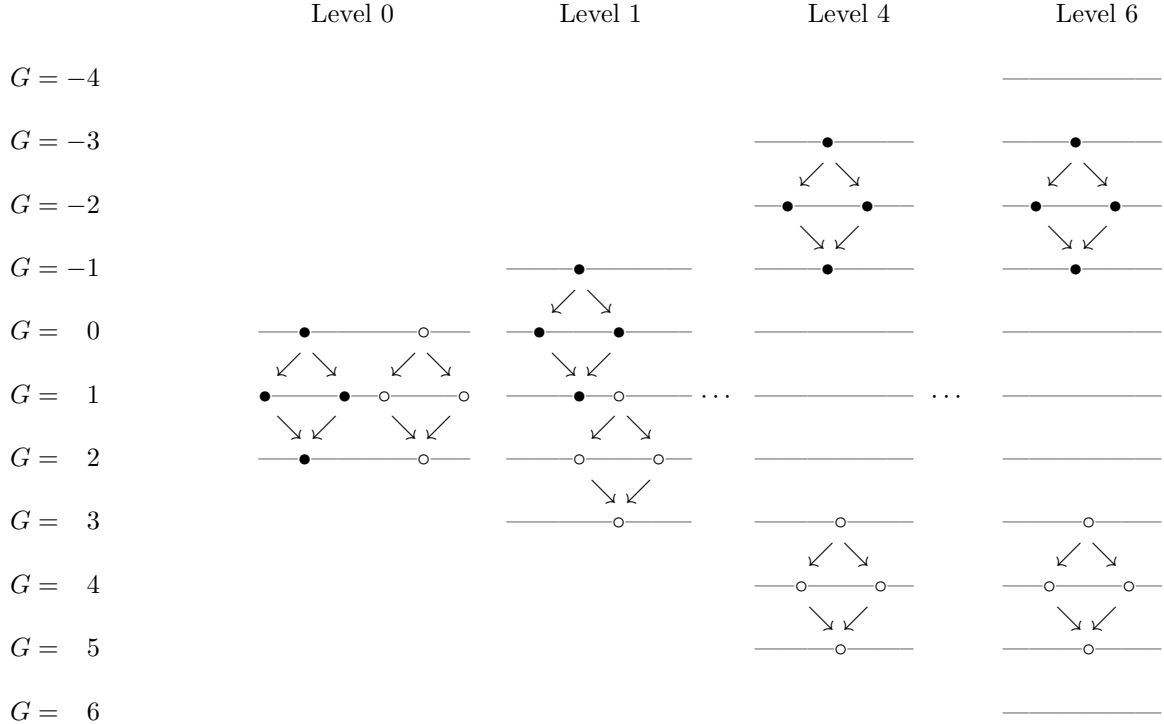
\centerline{\sl Figure 3}
{\tenpoint\sl This diagram shows the structure of the quartets and
their conjugates for each prime state, at levels 0, 1, 4 and 6 in the
two-scalar
$W_3$ string.  The notation is analogous to that of Fig.\ 2.  At level 0 there
are six tachyon prime states; at level 1 there are six prime states.  At level
4
there is one prime state, namely the $SL(2,C)$ vacuum.  At level 6 there are
two prime states, corresponding to the $x$ and $y$ operators described
above.}
\bigskip\bigskip

\bigskip
\noindent{\bf 4. Discussion}
\bigskip

     For the ordinary string in two dimensions, the ring of $G=0$ operators
generated by $x$ and $y$, together with their associated quartets and
conjugates, comprise the complete set of BRST-nontrivial discrete operators
(at least for the chiral sector).  For the two-scalar $W_3$ string, it is not
clear whether the states described in the previous section exhaust the BRST
non-trivial discrete states.  In fact the existence of the level-1 states
described above, which have no direct analogues in the ordinary string,
suggests that the story may be a more complicated one.  As we shall now
argue, they may indicate the existence of further $x$ and $y$-type
operators at higher level numbers.

     Recalling that we deduced the existence, and level number, for the $x$
and $y$ operators (3.11) by seeking operators that mapped tachyons into
tachyons, we can carry out a similar discussion for hypothetical operators
that map the set of level-1 prime states into itself.  In fact $x$ and $y$
themselves have this property: $x$ maps the level-1 state with momentum
$\vec p_6$ in (3.16) into the state with momentum $\vec p_5$, and $y$ maps
the state with momentum $\vec p_2$ into the state with momentum $\vec p_4$.
There are other pairs of level-1 states that satisfy the necessary condition
(3.9) for the existence of an operator that could map one into the other.
These would correspond to operators at level $\ell=8$ with momenta
$(\ft47Q_1,-\ft27Q_2)$, $(\ft47Q_1,-\ft{12}7Q_2)$, $(\ft17Q_1,Q_2)$ and
$(-\ft47Q_1,\ft{12}7Q_2)$; and operators at level $\ell=9$ with momenta
$(0,\ft{10}7Q_2)$, $(\ft57Q_1,-\ft57Q_2)$, $(\ft57Q_1,-\ft97Q_2)$ and
$(-\ft27Q_1,\ft{12}7Q_2)$.  These operators, having ghost number 0, would
correspond to states with ghost number $-3$.  The minimum allowed ghost
numbers at levels 8 and 9 are $-4$ and $-5$ respectively.  Thus to check
explicitly for the existence of such operators would be quite complicated,
and, especially in the $\ell=9$ case, would involve recognising and
discarding many BRST-trivial states.

     It is interesting to note that the six level-1 prime states, with momenta
given in (3.17), give rise to 12 physical states with the ``standard'' ghost
number, $G=0$.  Those with momenta $\vec p_5$ and $\vec p_6$ in (3.17)
correspond to physical states that have been found in earlier work on the
spectrum of the $W_3$ string [8,9].  However, the states with momenta
$\vec p_1,\ldots,\vec p_4$ are of a kind that have not been previously
described.  They correspond to physical states involving ghost, as well as
matter, excitations.  It would seem that the spectrum of physical states in
the $W_3$ string is thus richer than had previously been thought.

     The two-scalar realisation of $W_3$ is intimately related to the Lie
algebra of $SU(3)$.  It would not be surprising, therefore, if the discrete
states of the $W_3$ string were to exhibit an $SU(3)$ structure, possibly
with its enveloping algebra arising as a symmetry algebra of the states.
This would be a generalisation of the wedge subalgebra of $w_\infty$, found
as a symmetry of the states of the two-scalar ordinary string [3].

     In the two-scalar ordinary string, one can make an association between a
physical state $\ket{\chi}$ and a spin-1 current $j(z)$, defined by
$j(0)\ket{0}=b_{-1}\ket{\chi}$ [3].  The current $j(z)$ defines a charge that
automatically commutes with the BRST operator.  If one requires that $j(z)$ be
a
primary field, then the physical state $\ket{\chi}$ must be annihilated by
$b_0$.  It is these conserved currents that generate the symmetry algebra
described above.  For the two-scalar $W_3$ string, we can again build spin-1
currents from physical states, by acting with $b_{-1}$ on physical states.
These again give charges that commute with the BRST operator.  If we require
that the currents must be primary fields under the energy-momentum tensor,
then this imposes the restriction that the physical states must be
annihilated by $b_0$ and $\beta_0$.  In other words, these conditions on the
physical state $\ket{\chi}$ ensure that $b_{-1}\ket{\chi}$ is a
highest-weight state with respect to the Virasoro algebra.  In fact, they
also ensure that $b_{-1}\ket{\chi}$ is annihilated by the Laurent
modes $W_n$ of the spin-3 current, for $n>0$.   However, there seems to be no
way of having the state $b_{-1}\ket{\chi}$  also be an eigenstate of $W_0$.
Whether this presents a problem for  interpreting the conserved charges as
generators of a symmetry algebra is  not clear to us.  We note from (A.2-A.4)
in the appendix that for the $x$ and $y$ operators at level 6, the
corresponding states will indeed be annihilated by $b_0$, since $g_{11}$
vanishes, and that in each case there exists a choice of the parameters
$\lambda$ and $\tau$ such that the states are annihilated by $\beta_0$ as
well, by making the coefficient $g_{12}$ vanish.

     One might think that a three-scalar, rather than two-scalar,
realisation of $W_3$ would provide a more natural generalisation of the
two-scalar ordinary string.  In particular, the tachyons would then have
continuous momenta in the $\varphi_2$ and $\varphi_3$ directions (the
unfrozen ``spacetime'' dimensions in the terminology of [7,8,9]).  We have
looked at the conditions for the existence of level-6 $x$- and $y$-type
operators in this case, and found that there is now only one BRST
non-trivial example, with momentum $(\ft27 Q_1,0,0)$.  At level 1, we find
that of the six states given in (3.17), the last two generalise to discrete
states with momenta $(-\ft67Q_1,0,0)$ and $(-\ft87Q_1,0,0)$, whilst the
first four generalise to states with continuous ``spacetime'' momenta,
$(-\ft37Q_1,p_2,p_3)$ and $(-\ft47 Q_1,p_2,p_3)$.  In fact the phenomenon of
continuous-momentum states repeats at higher levels.  It can be understood
from the fact that the effective spactime theory for a $W$ string has fewer
gauge symmetries than an ordinary string, where two out of the spacetime
dimensions describe unphysical longitudinal modes.

\bigskip
\centerline{\bf Acknowledgments}
We are grateful to the International Centre for Theoretical Physics for
hospitality during the early stages of this work. C.N.P. would like to thank
Bengt Nilsson and Peter West for valuable discussions.   We made extensive
use of the Mathematica package OPEdefs for calculating operator-product
expansions [10].

\bigskip
\np
\centerline{\bf APPENDIX}
\bigskip
\bigskip
\bigskip

     In this appendix, we present the details of the discrete level-6
operators $x$ and $y$ in the two-scalar $W_3$ string.  They correspond to
states at ghost number $G=-3$, and can be written as
$$
R\, e^{\vec p\cdot\vec\varphi},\eqno(A.1)
$$
where the most general possible prefactor $R$ at this level and ghost number
takes the form
\bigskip
$$
\eqalign{
R&= g_1\ b\, b' \gamma \gamma' + g_2\ \beta c + g_3\ b\,\beta \gamma\gamma' +
g_4\ b' c +g_5\ \beta\gamma' +g_6\ b\, b' c\gamma\cr
&\ \ +g_7 b'\gamma' +g_8\ b''\gamma + g_9\ \beta'\gamma + g_{10}\ b \beta c
\gamma +g_{11}\ b\, c' +g_{12}\ b \gamma''\cr
&\ \ +g_{13}\ \varphi_1' b\gamma' + g_{14}\ \varphi_2' b\gamma' + g_{15}\
\varphi_1' b'\gamma + g_{16}\ \varphi_2' b' \gamma + g_{17}\ \varphi_1'
b\, c  +g_{18}\ \varphi_2' b\, c\cr
&\ \ +g_{19}\ \varphi_1'\beta\gamma + g_{20}\ \varphi_2' \beta\gamma +
g_{21}\ \varphi_1'' b\gamma + g_{22}\ \varphi_2'' b\gamma + g_{23}\
\varphi_1'' +g_{24}\ \varphi_2''\cr
&\ \ +g_{25}\ (\varphi_1')^2 +g_{26}\ \varphi_1'\varphi_2' + g_{27}\
(\varphi_2')^2 +g_{28}\ (\varphi_1')^2 b\gamma +g_{29}\ \varphi_1'\varphi_2'
b\gamma +g_{30}\ (\varphi_2')^2 b\gamma \ ,\cr}\eqno(A.2)
$$
with $'$ denoting the holomorphic derivative $\del$.

     As described in section 3, for general values of the momentum $\vec p$,
there is a unique solution, up to overall scale, for the coefficients $g_i$,
following from the physical-state requirement
$Q_{BRST}R e^{\vec p\cdot \vec\varphi(0)}\ket{0}=0$.
This corresponds to the BRST-trivial state
obtained by acting with $Q_{BRST}$ on the ghost-number $G=-4$ state
$b_{-1}b_{-2}\beta_{-1}\beta_{-2}\ket{\vec p,--}$.  At each of the two special
values of momentum $\vec p=(\ft27 Q_1,0)$ and $\vec p=(\ft17Q_1,\ft37 Q_2)$, a
second solution occurs; this is BRST non-trivial.  The two-parameter
family of solutions for  $\vec p=(\ft27 Q_1,0)$ is:
\bigskip
\bigskip
\settabs 3 \columns
\+$g_1={{4 i}\over{3 \sqrt{29}}}(63 \lambda-65\tau)\ ,$ &$g_2=-1566
\sqrt{2} (\lambda+\tau)\ ,$ &$g_3=-60\sqrt2 (\lambda+\tau)\ ,$\cr
\+$g_4=6 i \sqrt{29} (-9\lambda+7\tau)\ ,$&$g_5=18i\sqrt{29}(-7\lambda+
9\tau)\ ,$
&$g_6=12\sqrt2 (\lambda+\tau)\ ,$\cr
\+$g_7=-97\sqrt2(\lambda+\tau)\ ,$&$g_8=-32\sqrt2 (\lambda+\tau)\ ,$
&$g_9=48i\sqrt{29} (-3\lambda+\tau)\ ,$\cr
\+$g_{10}=72i\sqrt{29}(\lambda+\tau)\ ,$&$g_{11}=0\ ,$&$g_{12}=-8\sqrt2
(7\lambda+19\tau)\ ,$\cr
\+$g_{13}=-88\lambda+56\tau\ ,$&$g_{14}=0\ ,$&$g_{15}=8(\lambda+\tau)\ ,$\cr
\+$g_{16}=0\ ,$&$g_{17}=-24i\sqrt{58} (3\lambda+\tau)\ , $&$g_{18}=0\ ,
\qquad\qquad\qquad\qquad (A.3)$\cr
\+$g_{19}=-72i\sqrt{58}(\lambda+\tau)\ ,$&$g_{20}=0\ ,$&$g_{21}=32(2\lambda
+3\tau)\ ,$\cr
\+$g_{22}={224\over{\sqrt3}}\ \tau\ ,$&$g_{23}=24i\sqrt{58}(-4\lambda+
\tau)\ ,$
&$g_{24}=56i\sqrt{174}\ \tau\ ,$\cr
\+$g_{25}=-48i\sqrt{29}(2\lambda+\tau)\ ,$&$g_{26}=0\ ,$&$g_{27}=48i\sqrt{29}
\ \tau\ ,$\cr
\+$g_{28}=32{\sqrt 2}\lambda\ ,$&$g_{29}=0\ ,$&$g_{30}=32{\sqrt 2}\tau\ ,$\cr
\bigskip\bigskip
\noindent
where $\lambda$ and $\tau$ are arbitrary parameters. The BRST trivial
operator corresponds to the choice $\lambda=-\tau$. For the momentum
$\vec p=(\ft17Q_1,\ft37 Q_2)$, the two-parameter family of solutions is:
\bigskip\bigskip
\settabs 3 \columns
\+$g_1={{4 i}\over{3 \sqrt{29}}}(129 \lambda-127\tau)\ ,$ &$g_2=-1566
\sqrt{2} (\lambda+\tau)\ ,$ &$g_3=-60\sqrt2 (\lambda+\tau)\ ,$\cr
\+$g_4=6 i \sqrt{29} (-15\lambda+17\tau)\ ,$&$g_5=18i\sqrt{29}(-17\lambda+
15\tau)\ ,$
&$g_6=12\sqrt2 (\lambda+\tau)\ ,$\cr
\+$g_7=-97\sqrt2(\lambda+\tau)\ ,$&$g_8=-32\sqrt2 (\lambda+\tau)\ ,$
&$g_9=48i\sqrt{29} (-3\lambda+5\tau)\ ,$\cr
\+$g_{10}=-72i\sqrt{29}(\lambda+\tau)\ ,$&$g_{11}=0\ ,$&$g_{12}=-8\sqrt2
(25\lambda+\tau)\ ,$\cr
\+$g_{13}=64\lambda-80\tau\ ,$&$g_{14}=16{\sqrt 3} (4\lambda-5\tau)\ ,$
&$g_{15}=4(\lambda+\tau)\ ,$\cr
\+$g_{16}=4{\sqrt 3}(\lambda+\tau)\ ,$&$g_{17}=48i\sqrt{58}\tau\ , $
&$g_{18}=48i\sqrt{174}\tau\ ,\qquad\qquad (A.4)$\cr
\+$g_{19}=36i\sqrt{58}(\lambda+\tau)\ ,$&$g_{20}=36i\sqrt{174}(\lambda+\tau)
\ ,$&$g_{21}=32(7\lambda-\tau)\ ,$\cr
\+$g_{22}={128\over{\sqrt3}}\ \tau\ ,$&$g_{23}=24i\sqrt{58}(-7\lambda+
5\tau)\ ,$ &$g_{24}=64i\sqrt{174}\ \tau\ ,$\cr
\+$g_{25}=48i\sqrt{29}(-\lambda+\tau)\ ,$&$g_{26}=48i\sqrt{87}
(\lambda+\tau) ,$&$g_{27}=96i\sqrt{29}\ \tau\ ,$\cr
\+$g_{28}=32{\sqrt 2}\lambda\ ,$&$g_{29}=32{\sqrt 6}(-\lambda+\tau)\ ,$
&$g_{30}=32{\sqrt 2}\tau\ .$\cr
\bigskip\bigskip
\noindent
The BRST-trivial solution again corresponds to $\lambda=-\tau$.  Note that
for both cases, the coefficient $g_{11}$ vanishes.  From $(A.2)$, we see
that this implies that the corresponding states $x(0)\ket{0}$ and
$y(0)\ket{0}$ are annihilated by $b_0$.   By adjusting the parameters
so that the coefficient $g_{12}$ vanishes too, we obtain states that are
annihilated by $\beta_0$ as well.  As discussed in section 4, these states
may be interpreted as giving rise to conserved currents.

\vskip 3truecm
\centerline{\bf REFERENCES}
\bigskip
\item{1.}A.M. Polyakov, {\sl Mod. Phys. Lett.} {\bf A6} (1991) 635.
\item{2.}B. Lian and G. Zuckerman, {\sl Phys. Lett.} {\bf B254} (1991) 417.
\item{3.}E. Witten, {\sl Nucl. Phys.} {\bf B373} (1992) 187.
\item{4.}E. Witten and B. Zwiebach, {\sl Nucl. Phys.} {\bf B377}
(1992) 55.
\item{5.}V.A. Fateev and A.B. Zamolodchikov, {\sl Nucl. Phys.} {\bf B280}
(1987) 644.
\item{6.}J. Thierry-Mieg, {\sl Phys. Lett.} {\bf B197} (1987) 368.
\item{7.}C.N. Pope, L.J. Romans and K.S. Stelle, {\sl Phys. Lett.} {\bf B268}
(1991) 167.
\item{8.}C.N. Pope, L.J. Romans, E. Sezgin and K.S. Stelle, {\sl Phys.
Lett.} {\bf B274} (1992) 298.
\item{9.}H. Lu, C.N. Pope, S. Schrans and K.W. Xu, preprint, CTP-TAMU-5/92,
to appear in {\sl Nucl. Phys.} {\bf B}.
\item{10.}K. Thielemans, {\sl Int. J. Mod. Phys.} {\bf C2} (1992) 787.
\end